\documentclass[aps,twocolumn,nofootinbib,preprintnumbers,superscriptaddress,floatfix]{revtex4-2}
\usepackage{graphicx}
\usepackage{color}
\usepackage{bm}
\usepackage{amsmath}
\usepackage{amssymb}
\usepackage{xspace}
\usepackage[normalem]{ulem}
\usepackage{units}
\usepackage{dcolumn}
\usepackage{paralist}
\usepackage{natmove}
\usepackage{natbib}
\usepackage[table]{xcolor}
\usepackage{upgreek}

\usepackage{hyperref}

\begin{document}

\title{FluxGAN: A Physics-Aware Generative Adversarial Network Model for Generating Microstructures That Maintain Target Heat Flux}
\author{Artem~K.~Pimachev}
\affiliation{Ann and H.J. Smead Aerospace Engineering Sciences, University of Colorado Boulder, Boulder, CO 80303, USA}
\author{Manoj~Settipalli}
\affiliation{Ann and H.J. Smead Aerospace Engineering Sciences, University of Colorado Boulder, Boulder, CO 80303, USA}
\author{Sanghamitra~Neogi}
\email{sanghamitra.neogi@colorado.edu}
\affiliation{Ann and H.J. Smead Aerospace Engineering Sciences, University of Colorado Boulder, Boulder, CO 80303, USA}

\date{\today}

\begin{abstract}
We propose a physics-aware generative adversarial network model, FluxGAN, capable of simultaneously generating high-quality images of large microstructures and description of their thermal properties. During the training phase, the model learns about the relationship between the local structural features and the physical processes, such as the heat flux in the microstructures, due to external temperature gradients. Once trained, the model generates new structural and associated heat flux environments, bypassing the computationally expensive modeling. Our model provides a cost effective and efficient approach over conventional modeling techniques, such as the finite element method (FEM), for describing the thermal properties of microstructures. The conventional approach requires 
computational modeling that scales with the size of the microstructure model, therefore limiting the simulation to a given size, resolution, and complexity of the model. In contrast, the FluxGAN model uses synthesis-by-part approach and generates arbitrary large size images at low computational cost. We demonstrate that the model can be utilized to generate designs of thermal sprayed coatings that satisfies target thermal properties. Furthermore, the model is capable of generating coating microstructures and physical processes in three-dimensional (3D) domain after being trained on two-dimensional (2D) examples. Our approach has the potential to transform the design and optimization of thermal sprayed coatings for various applications, including high-temperature and long-duration operation of gas turbines for aircraft or ground-based power generators.
\end{abstract}

\maketitle

\section*{Introduction}

In recent years, machine learning techniques, such as generative adversarial networks (GANs), are increasingly being used to explain deterministic multiphysics phenomena where the underlying physical model is complex or unknown. Physics-informed generative adversarial networks have been proposed for solving forward, inverse, and mixed stochastic problems using data from a limited number of scattered measurements~\cite{yang2020physics}. Instead of relying solely on data for training, this approach encoded the governing physical laws in the GAN architecture in the form of stochastic differential equations. FluidGAN has been proposed for learning and predicting time-dependent convective flow coupled with energy transport~\cite{jiang2020deep}. StressGAN model was developed to generate mechanical stress distributions for various complex cases of geometry, load, and boundary conditions~\cite{jiang2021stressgan}. The model was shown to predict more accurate high-resolution stress distributions than a convolutional neural network model. In parallel, GANs have shown remarkable progress for creating high-resolution synthetic images, visually similar to their real counterparts, supported by revolutionary advents in the graphical data domain. Until recently, training such model required enormous amount of training images and was highly unstable. GANs have been shown to be highly effective in generating realistic microstructural data that can be used to augment limited experimental data or simulate microstructures that are challenging to obtain experimentally~\cite{hsu2021microstructure,nguyen2022synthesizing}. The style-based GAN (StyleGAN2) architecture~\cite{karras2019style}, implemented the style transfer approach~\cite{huang2017arbitrary} with the adaptive discriminator augmentation (ADA)~\cite{karras2020training} mechanism, and demonstrated that high-quality images can be generated with limited training data. 

In this study, we propose a new FluxGAN model that simultaneously generates high-quality microstructural images and high-resolution description of thermal properties of these microstructures. We adopt the StyleGAN2 architecture to develop the model and train the model using images that include both microstructural information and the associated physical (i.e., thermal) phenomena. We demonstrate the capabilities of our approach for the analysis and design of microstructure-controlled thermal properties of thermal sprayed coatings. The thermal sprayed coatings exhibit highly diverse structural environments, including pores with different shapes and sizes, and varied size grains aligned at different crystallographic orientations. The microstructural features strongly affect interface-sensitive material properties, such as thermal diffusivity. Additionally, the structural features could change and evolve due to sintering in high temperature environments. The presence of microstructural features is often quantified by the porosity parameter: (pore volume)/(total volume)$\times 100$ \%.
The variation of thermal properties of coatings has commonly been described in relation with porosity, using phenomenological models~\cite{wang2006new}. However, global parameter (e.g. porosity) based models may not accurately capture the physics of heat transfer across various microstructural features and feature interfaces. The complexity of structural features and their evolution makes it highly challenging to characterize the physical properties of these coatings, especially after aging, using existing experimental and numerical approaches.

Our proposed FluxGAN model provides highly accurate prediction of thermal phenomena in coatings with different microstructures that include different porosities, grain sizes, and crystallographic orientations. We use a small number of two-dimensional (2D) images of niobium thermal sprayed coatings, obtained with backscatter electron (BSE) imaging technique, as the `true' image of the coating microstructures. We compute the thermal properties of the microstructures using the image-based finite element method (FEM). We train the FluxGAN model using superimposed images that include both microstructural and thermal property data. The FluxGAN model combines the physical properties predicted by FEM and the GAN architecture, for generating diverse microstructures with desired thermal properties. We use separate FEM simulations to validate the generated thermal data. The model uses synthesis-by-part approach and generates images of arbitrary size patch by patch, which requires constant memory usage. In contrast, the FEM memory usage and computational time scales with the image size. The combination of image-based FEM and GAN models offers a promising solution to overcome the limitations of traditional experimental and numerical methods. Our model provides a cost effective and efficient approach for describing the microstructure-controlled thermal properties of thermal sprayed coatings. The model establishes a direct relationship between structural features and thermal properties of coatings, which could be utilized to better control and optimize their thermal properties. We demonstrate that our model can be used to design thermal sprayed coatings that can maintain tailored heat flux that meets application-specific requirements. Our approach has the potential to transform the design and optimization of thermal sprayed coatings for various industrial applications, including high-temperature and long-duration operation of gas turbines for aircraft or ground-based power generators.

\begin{figure*}
\begin{center}
\includegraphics[width=0.9\linewidth]{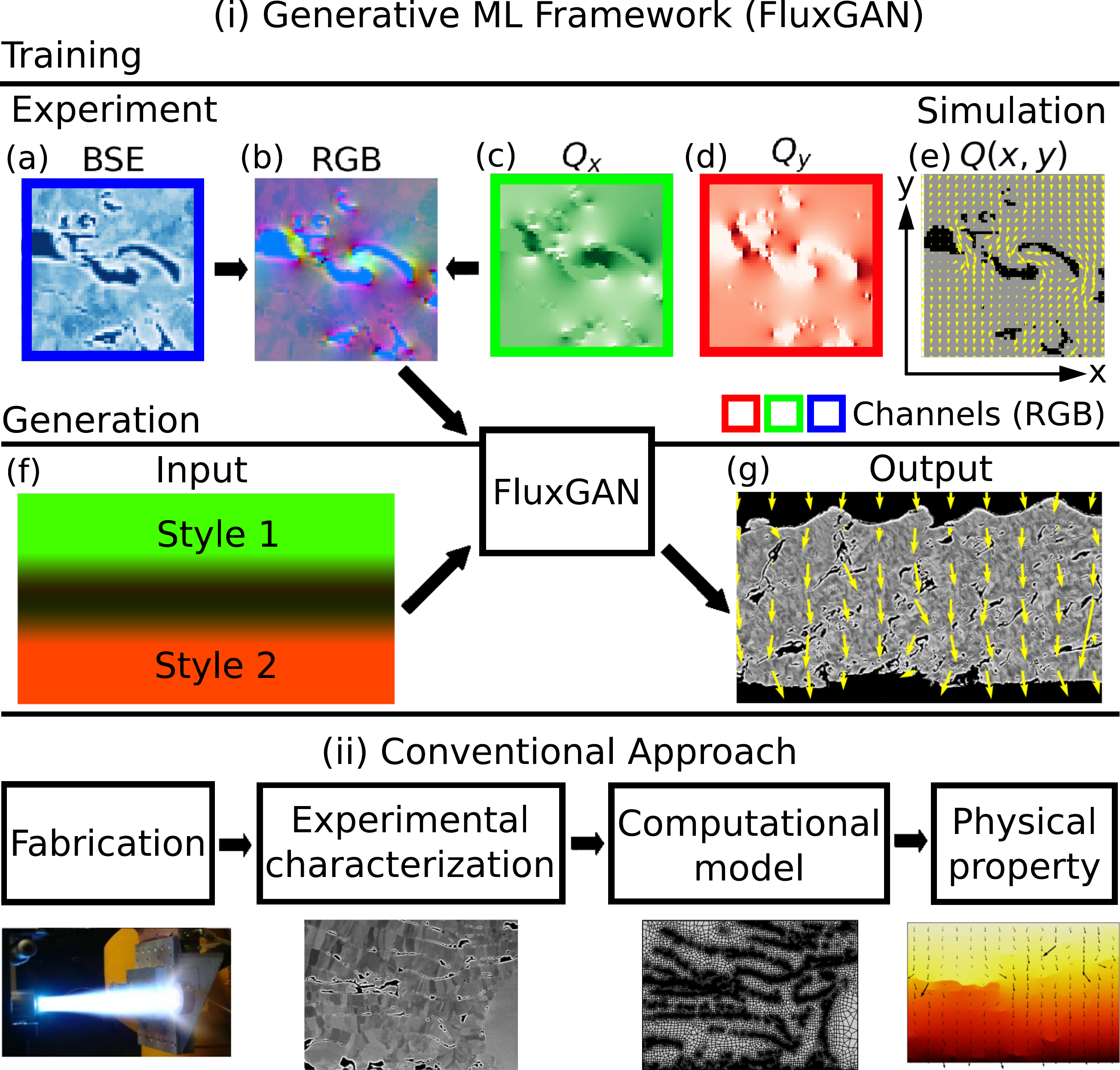}
\caption{{\bf Overview of FluxGAN model and its advantage over conventional approach:} (i) The model includes a training and a generation phase. We train the model with (b) superimposed RGB images that include (a) experimental structural images and (c,d) heat flux data, simulated with FEM techniques. The model classifies the distinct features of the RGB images and assigns them as different styles. Once trained, the model generates images based on user input. (f) Spatial map of different styles as input. (g) The model generates new structural image superimposed with associated heat flux, that satisfies target thermal properties. (ii) In contrast, conventional design approach follows an expensive and iterative cycle until microstructures with target physical properties are obtained.}
\label{fig:workflow}
\end{center}
\end{figure*}

\section*{Methods}

\subsection*{Model Overview}

Figure~\ref{fig:workflow}(i) shows the overview of our physics-informed FluxGAN model. The model includes a training and a generation phase. We train the model with images that include information about structural as well as thermal properties of the microstructure. We obtain the 2D structural images, as shown in Fig.~\ref{fig:workflow}(a), using backscatter electron (BSE) imaging and simulate the thermal properties using FEM techniques. Figure~\ref{fig:workflow}(e) shows the spatial distribution of heat flux, $Q(x,y)$, generated in the microstructure, due to the application of temperature gradient along the top-to-bottom direction. Figure~\ref{fig:workflow}(c) and (d) show the graphical domain representation of the heat flux components, $Q_x$ and $Q_y$, respectively. We create RGB images by superimposing the experimental structural images and the heat flux maps, as shown in Fig.~\ref{fig:workflow}(b). We provide the RGB images as input to the FluxGAN model. The model identifies the different features of the RGB training images, referred to as styles, in an unsupervised manner. The styles not only classify structural features but are strongly associated with heat flux patterns of the RGB images. In the unguided 
generation phase, the FluxGAN model requires minimal input from the user and generates a new image corresponding to the input. An example input is shown in Fig.~\ref{fig:workflow}(f). The input specifies the overall height and width of the image and the spatial map of styles to be included. The spatial map includes the number of styles that can be identified in the image (styles 1 and 2), the locations for regions with styles 1 and 2 (anchors), and the merging function that can be applied to obtain smooth interfaces between the regions. Figure~\ref{fig:workflow}(g) shows a representative microstructure, generated by the FluxGAN model. The generated image includes structural features that correspond to styles 1 and 2 as specified by the user input. The image also includes information about the heat flux that will be observed if the microstructure is subjected to temperature gradients along the top-to-bottom image direction. Thus, the FluxGAN model directly generates designs of representative microstructures and corresponding physical properties from a minimal user input. The model can even generate designs of new classes of structures that are not present in original experimental images. For example, different designs can be created by combining different styles, varying the positions of the styles in two-dimensional (2D) or three-dimensional (3D) spatial domains. Additionally, the model offers guided assignment of styles in the generated microstructures, based on the specifications of the input image. For example, given a known heat flux distribution in the heat source or sink region as input, the model allows for inverse design of microstructures that satisfies the target heat flux condition. We describe the different components of the FluxGAN model in the following sections. We demonstrate the effectiveness of the FluxGAN model by using single phase niobium (Nb) thermal spray coatings as an example. However, our model can be broadly applied to different materials and structures to assess how well they satisfy target physical properties.  

The conventional approach for designing new coatings, as outlined in Fig.~\ref{fig:workflow}(ii), typically iterates through the following steps: (1) fabrication of coatings, (2) experimental characterization, (3) developing computational models based on experimental images, (4) physical property calculation, and repeat (1-4) till the fabricated structures satisfy target physical properties. Our FluxGAN model provides an expedited approach to inform the design of new microstructures that satisfy target physical properties, bypassing the expensive iterative process. In this study, we have chosen the styles in an arbitrary manner. However, the styles can be associated with specific physical conditions. For example, the different spray intervals in a thermal spray process could result in different structural features. The different features can also be associated with diverse environmental conditions, such as high temperatures or corrosion. As styles are strongly associated with the heat flux, physical meaning of styles can be regions of high or low thermal conductance and temperature hot spots, etc. A FluxGAN model aware of the relationship between the styles and synthesis or environmental conditions can offer great practical advantage. For example, the styles in generated images can inform strategies to synthesize microstructures with target functionalities and also, provide information about the reliability of microstructure in diverse environmental conditions.

\section*{Training Phase}
\subsection*{Processing of Experimental Images}

\begin{figure}
\begin{center}
\includegraphics[width=0.9\linewidth]{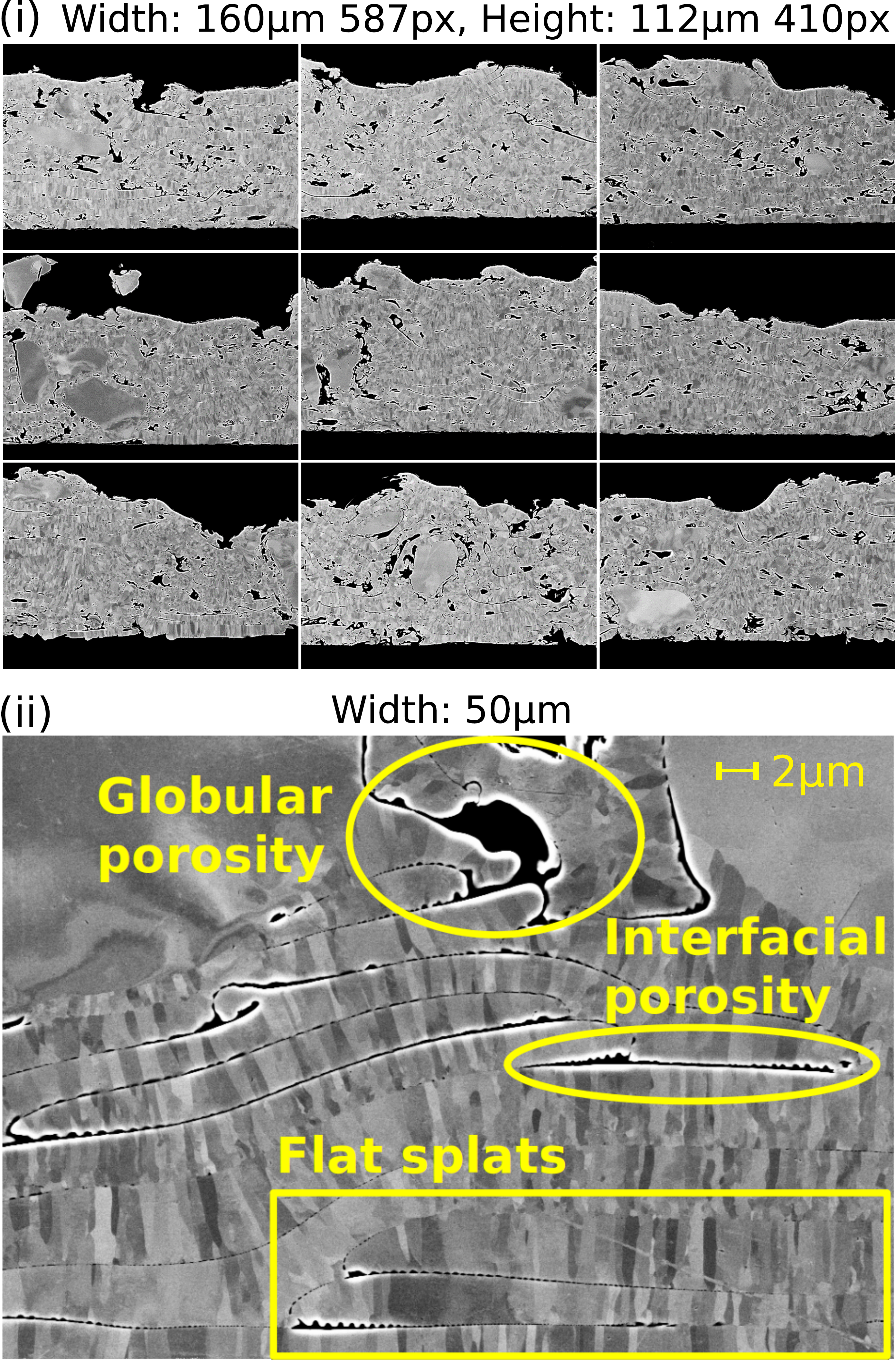}
\caption{{\bf Experimental images of coating microstructures:} (i) Raw BSE images of niobium thermal spray coatings. (ii) Magnified image of microstructure shows a wide variety of structural features, such as flat splats and porosities due to conditions during the synthesis process.}
\label{fig:expimages}
\end{center}
\end{figure}

Figure~\ref{fig:expimages} shows the 2D experimental BSE images of the Nb thermal spray coatings that we use to demonstrate the performance of the FluxGAN model. The dimension of each image is 112$\upmu$m $\times$ 160$\upmu$m (410px $\times$ 587px). Three distinct regions can be visually identified from the images: the gray middle region that shows the presence of the coating microstructure, the top and bottom dark regions where no coating material is present and the two interface regions between the gray coating and the dark regions. We assume that the dark regions of the images represent presence of air. We show a magnified version of a representative coating microstructure in Fig.~\ref{fig:expimages}(ii). As can be noted, the microstructure includes distinct features, such as flat splats and globular and interfacial porosities.
In addition to these features, a wide variety of small features can also be noted that are representative of materials properties or processing conditions. However, some features may appear due to difference in lighting conditions during imaging. Note that the FluxGAN model identifies the different styles of the training images in an unsupervised manner. To ensure that the microstructure images only include `true' features, we preprocess the grayscale BSE images and remove features that appear due to irregular image brightness. The preprocessing is an essential step since our goal is to develop a model that can relate `true' structural features with the corresponding physical properties. First, we obtain the pixel intensity distribution of the BSE images, by counting the number of pixels that has a given intensity in the range between 0 and 255. Figure~\ref{fig:expimages}(a) shows that the coating region of the images have intensities $\ge$ 50 in this scale. The appearance of the distinct peaks can be attributed to different lighting conditions during imaging. We scale the intensities to values in the range between 0 and 1. We then shift the mean of the distribution to 0.5 using a power transformation. Finally, we rescale the pixel intensity distribution of the coating region to be in the range from 50 to 255. Figure~\ref{fig:expimages}(b) shows the pixel intensity distribution after the transformation. The intensity values below 50 represent pores or air regions where the coating material is not present.  

\subsection*{Augmentation of Training Set}

\begin{figure}
\begin{center}
\includegraphics[width=0.9\linewidth]{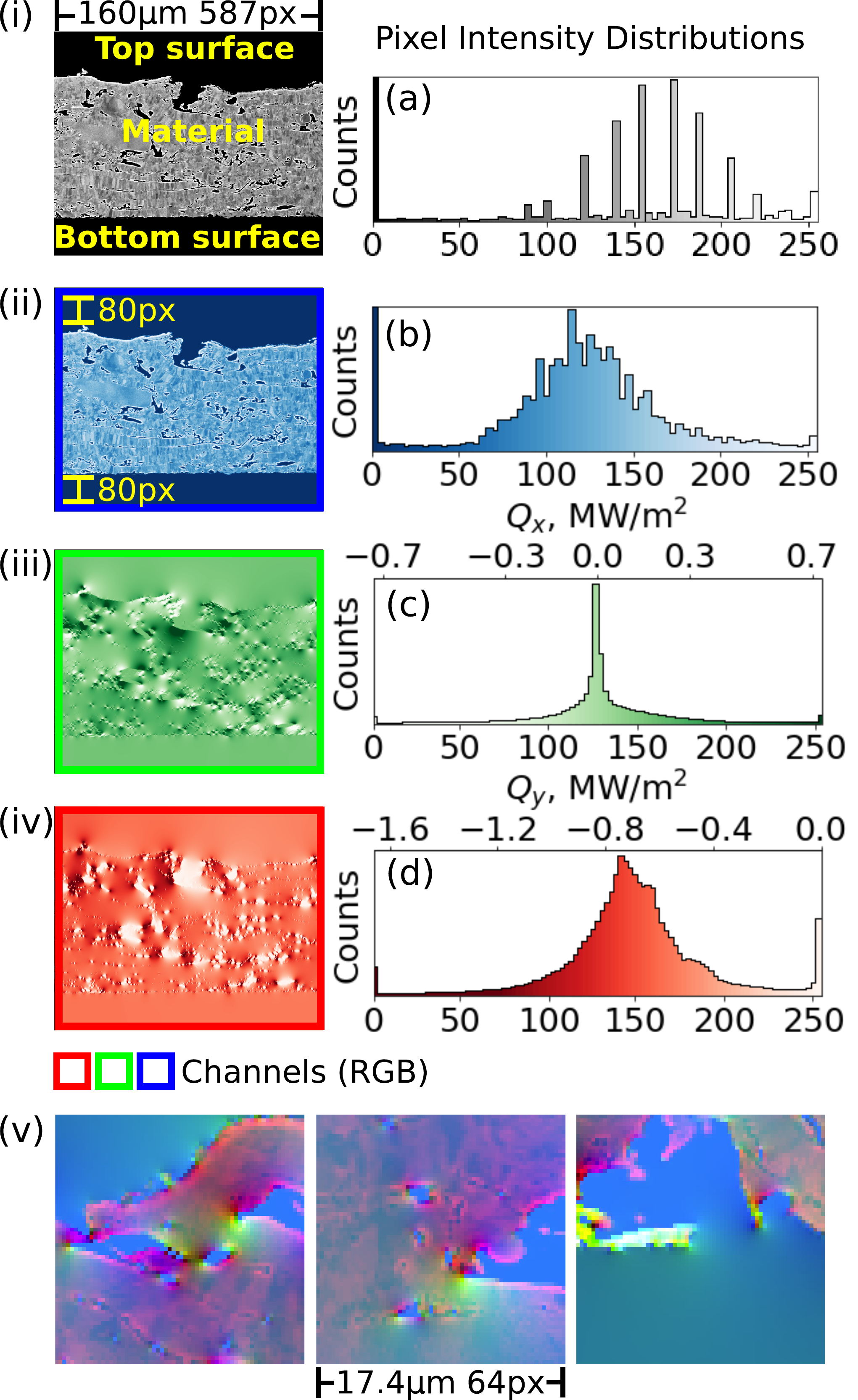}
\caption{{\bf Preparation of training images:} (i) BSE image. Pixel intensity distribution of BSE images (a) before and (b) after image processing;  Histogram shows the number of pixel counts for different pixel intensity in a monochrome channel. Heat flux along components, (c) $Q_x$,  and (d) $Q_y$, for applied temperature gradients. Top horizontal axes labels show heat flux values and the bottom labels show graphical domain encoding for the monochrome channels. Three channel representations of (ii) structural and (iii-iv) heat flux data. The bottom label values of (b-d) are used as color scheme of images (ii-iv). The three channels are superimposed to create RGB images. (v) Tiles extracted from RGB images are used as training images.}
\label{fig:distributions}
\end{center}
\end{figure}

Our experimental dataset is limited to nine BSE images, as shown in Fig.~\ref{fig:expimages}. However, the GAN models require enormous amount of training images to generate high-resolution images as output. Extensive efforts are ongoing to develop models that can create high quality images with limited training data~\cite{gurumurthy2017deligan,gulrajani2017improved}. Recent approaches, such as StyleGAN2~\cite{karras2019style}, require a smaller training set, however, still require substantial number of training images. We implement different strategies to augment the training dataset. We apply rotation and reflection operations on the raw images and add the new images to the dataset. The new images include additional variations of structural features that impact the thermal properties. We thus obtain a diverse set of images to train the FluxGAN model. However, it is essential that the new images include features that are consistent with the physical conditions displayed by the original images. This constraint imposes restrictions on the type of operations we could perform to create new images. For example, all experimental images of Fig.~\ref{fig:expimages} (and also, Fig.~\ref{fig:distributions}(i)) show that the images have flatter bottom surfaces and uneven top surfaces. These features appear because the thermal spray coatings are fabricated on top of substrates. We only apply horizontal flip transformations but do not apply any vertical flip operations to maintain similar layer-like features in the new images. Additionally, when the droplets of melted material are sprayed in a regular interval, the material solidifies and forms layers in the top-to-bottom direction. As can be seen from Fig.~\ref{fig:expimages}(ii), the layers are highly uneven. This is because the sprayed material might solidify at different rates in the sprayed microstructures during fabrication. Highly anisotropic features may appear when droplets of melted material are sprayed over already solidified splats. To ensure that the new images retain such characteristics, we apply only small angular rotations between -2.5 to +2.5 degrees, with 11 incremental steps. We set the top and bottom air regions to be 80 pixels wide and add extra space as needed so that the coating region stays within the image boundary, as shown in Fig.~\ref{fig:distributions}(ii). The augmented dataset contains a total of 198 (9 original $\times$ 11 rotated $\times$ 2 flipped) superimposed images.

\subsection*{Computation of Thermal Properties}

We compute the thermal properties of the coating microstructures using image-based FEM techniques, as implemented in COMSOL software~\cite{multiphysics1998introduction}. We assume that the heat conduction direction is along the top-to-bottom direction of the coatings. We convert the microstructure images to masks by assigning all pixels with intensities $<$ 50 to black color, representing air, and the pixels with intensities $\ge$ 50 to grey color, representing niobium. Figure~\ref{fig:workflow}(a) and (e) show an experimental image and the corresponding masked image, respectively. We use the masked images as input for image-based FEM simulations. We import the masked images to COMSOL and use the built-in function, $im(x,y)=\{0, 1\}$, to mark different regions within each image. The discrete values correspond to black and gray regions of the images, respectively. This creates rectangular simulation domains that have same dimensions as the input images. We use the image function, $im(x,y)$, and the thermal conductivity values of air and Nb, to define spatially resolved thermal conductivities, $\kappa_\text{local}(x,y)$, in the rectangular simulation domain:
\begin{equation}
    \kappa_\text{local}(x,y) = 
    \begin{cases}
        \kappa_\text{Air},& \text{if }im(x,y)= 0\\
        \kappa_\text{Nb},              &\text{if }im(x,y)= 1
    \end{cases}
    \label{eq:KappaLocal}
\end{equation}
We choose thermal conductivity of niobium ($\kappa_\text{Nb}=54$ Wm$^{-1}$K$^{-1}$) for all grey pixels and thermal conductivity of air ($\kappa_\text{Air}=0.1$ Wm$^{-1}$K$^{-1}$) for all black pixels. 
We also define the local densities, $\rho_\text{local}(x,y)$, and local specific heat capacities, $C_\text{p, local}(x,y)$, using similar equations as Eq.~\ref{eq:KappaLocal}. We choose $\rho_\text{Air,Nb}=\{1.225, 8570\}$ Kgm$^\text{-3}$ and $C_\text{p}^{\text{Air, Nb}}=\{1008, 260\}$ JKg$^\text{-1}$K$^\text{-1}$, respectively. However, the heat flux distribution is primarily determined by the local thermal conductivities, as expected. The thermal properties of porous materials are commonly calculated using effective models, such as the Maxwell-Eucken model~\cite{wang2006new}. These models only take the porosity of the structure into account and do not consider the distribution of pores and their shapes to calculate thermal properties. Instead, we compute the heat flux in the coatings using image-based FEM and Fourier's law: 
$\textbf{q} = -d_z\kappa_\text{local}\nabla T$.
Here, $q$ is the heat flux along the temperature gradient $\nabla T$ and $d_z$=1m is the thickness along $z$ direction. We only discuss the heat flux in the 2D $x-y$ plane of the coatings and report the values of $Q=q/d_z$, in units of Wm$^{-2}$. 
We find that the heat flux and the resulting effective thermal conductivity can be significantly different for two coatings with the same porosity but different pore distributions. Such variability cannot be captured by the effective models.

We assume that both the top and the bottom no-material regions are made of low thermal conductivity material, i.e., air. We model the top region as heat source with temperature equal to 700K and the bottom region as heat sink with temperature 300K. Thus, we simulate an applied temperature bias of 400 K along the top-to-bottom direction of the coatings, for all BSE masked images. We model the lateral boundaries of the simulation cells as perfect insulators. We define a rectangular grid for the FEM simulations of dimension, $p_y \times p_x$. Here $p_x$ and $p_y$ are the number of pixels along the $x$ and the $y$ directions of the images. We use an adaptive meshing scheme and carry out five refinements in COMSOL to identify a mesh grid that minimizes the $L_2$ norm of the squared error. The image function $im$ is defined on the grid and assigns the local material properties accordingly. We compute heat flux, $q$, using these adaptive nodes and element shape functions. For each image, we iteratively solve for $q$ that satisfies the boundary conditions and local thermal conductivities. We assume d$_z$ = 1m in our calculations, thus, $Q$ have the same values as $q$, but different units. We refer to $Q(x,y)$ as heat flux in the subsequent discussion. $Q(x,y)$'s are defined on a rectangular grid of the same resolution as the input masked image. However, the grid is different for each image, depending on the original resolution of the experimental image and the applied transformations as described in the `Augmentation of Training Set' section. The associated heat flux also vary significantly, as can be expected. 

We show the components of heat flux along $x$ and $y$ directions in Fig.~\ref{fig:distributions}(c,d), respectively. The labels of the top horizontal axes show the heat flux values. We transform $Q_x$ and $Q_y$ to represent them in the graphical domain.
We linearly scale $Q_x$ and $Q_y$ such that the middle 98\% of the distribution falls in the range from 0 to 255. We assign the left and the right 1\% of heat flux distribution to values of 0 and 255, respectively. We transform the $Q$ values using the following equations: $Q_x=Q_x/10^6\times 173 + 127$; $Q_y=Q_y/10^6 \times 153 + 255$. We obtain these equations by assigning the bounds of the middle 98\% of the $Q$-distributions to graphical encoding values. 
The bottom axes labels of Fig.~\ref{fig:distributions} (c,d) correspond to the graphical domain encoding for the monochrome channels.
Note that the FEM simulation domains have horizontal reflection symmetry. We leverage this aspect and obtain results for 99 additional images by horizontally reflecting the FEM-generated heat flux $Q (x,y)$ [W/m$^2$]. In the reflected images, the ($x,y$) values are flipped horizontally and the magnitudes of $Q_x$ change sign, however, the magnitudes of $Q_y$ stays the same. Thus, we obtain data for 198 images from the 99 FEM calculations.

\begin{figure}
\begin{center}
\includegraphics[width=0.8\linewidth]{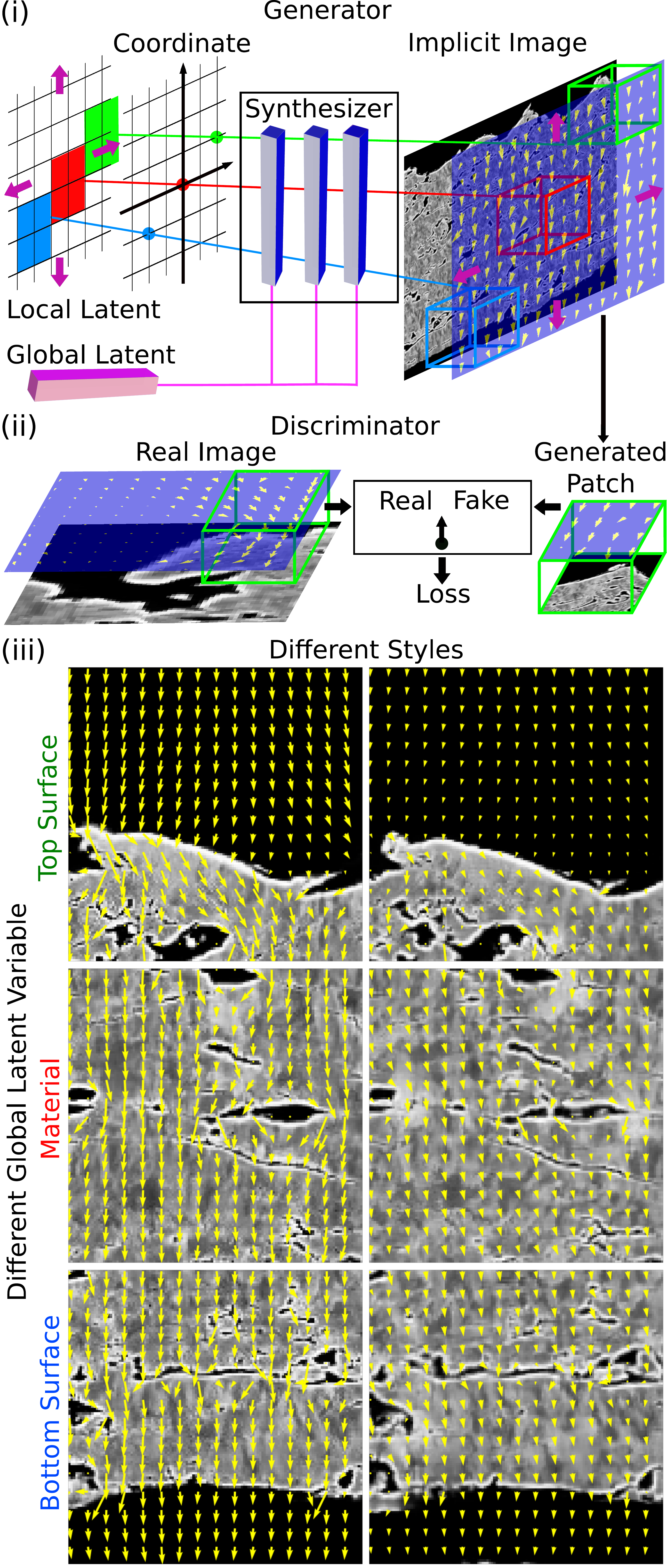}
\caption{{\bf Schematic of the FluxGAN model workflow.} Model architecture consists of (i) a generator, and (ii) a discriminator. (iii) Large-size images are generated by learning spatially extensible representations living in a latent space. Images in the first and the second columns are produced by varying global latent and styles respectively. Top, middle, and bottom images correspond to the green, red, and blue boxed regions in the implicit image (i), respectively.}
\label{fig:diversity}
\end{center}
\end{figure}

\subsection*{Preparation of Training Images}

We superimpose structural images and the corresponding heat flux maps into three-channel RGB images. The R, G, and B channels are the graphical encoding of $Q_y$, $Q_x$, and the structural images, as shown in Fig.~\ref{fig:distributions}(d), (c) and (b), respectively. We train the FluxGAN model on the relationship between the structural features of the coatings and the associated thermal properties. To this end, we create a training set by uniformly sampling overlapping square tiles of size 17.4$\upmu$m $\times$ 17.4$\upmu$m (64px $\times$ 64px) from the RGB images. We split each RGB image into $19 \times 13$ overlapping tiles and create a training set that includes 48,906 ($=198\times 19 \times 13$) RGB images. Figure~\ref{fig:distributions}(v) shows representative square tiles included in the training set. Each image is represented by an array of numbers. The dimension of the array is the height times the width of the tile, times the number of channels in the image: $64 \text{px} \times 64 \text{px} \times 3$. The array elements are pixel intensities in the range from 0 to 255, that correspond to the graphical encoding of the heat flux maps and the structural images. 

\subsection*{Unsupervised Learning of Styles}

We train the FluxGAN model using square 64px $\times$ 64px RGB patches, as discussed in the previous section. GAN models can learn from small patches and successfully generate new large-size images. However, the reconstruction of large scale holistic images can be challenging. For example, techniques such as consecutive image outpainting can result in discontinuous and repetitive structures~\cite{abdal2020image2stylegan++,liu2021infinite}. Several models have been proposed recently that can generate large images but they fail to maintain a global coherence in the generated images~\cite{shaham2019singan,shocher2019ingan,lin2019coco}. The recently proposed InfinityGAN approach has shown a superior ability to generate holistic large-size images after being trained on small patches~\cite{lin2021infinitygan}. The generated images show a great diversity and continuity along all spatial directions. We implement the InfinityGAN approach in our FluxGAN model to generate large coating images, learning from small patches. The approach generates large holistic images, also known as implicit images, with diverse local features. The InfinityGAN architecture adopts functionality from StyleGAN2~\cite{karras2019style}, which the FluxGAN model leverages to learn the different styles or the textures of the images, in an unsupervised manner. InfinityGAN synthesizes images of arbitrarily large-size by learning spatially extensible representations living in a latent space. The latent space is encoded by a global latent vector that guides the holistic layout of the infinite-pixel image, and a local latent tensor that expresses structural variations of small region patches. 


We show a schematic of the FluxGAN model workflow in Fig.~\ref{fig:diversity}. The model consists of a generator that learns to generate plausible images (Fig.~\ref{fig:diversity}(i)), and a discriminator that learns to distinguish the generated fake images from real images (Fig.~\ref{fig:diversity}(ii)). The generator consists of two modules, a structure and a texture synthesizer. The structure synthesizer is a neural implicit function~\cite{park2019deepsdf,mescheder2019occupancy,mildenhall2021nerf,chen2021learning}, that takes three sets of inputs: a global latent variable, a local latent variable and a continuous coordinate that samples the sub-regions of the implicit image. The texture synthesizer is a fully convolutional StyleGAN2 model~\cite{karras2019style}, with all positional information removed. In addition to the same three inputs, a set of randomized noises are provided as input to the texture synthesizer, to model variations of fine-grained texture within the local structural representations. 
A mapping layer projects global latent onto a texture style vector, and the texture style is injected into all pixels in each layer of the global image via feature modulation. 
\begin{figure}
\begin{center}
\includegraphics[width=0.9\linewidth]{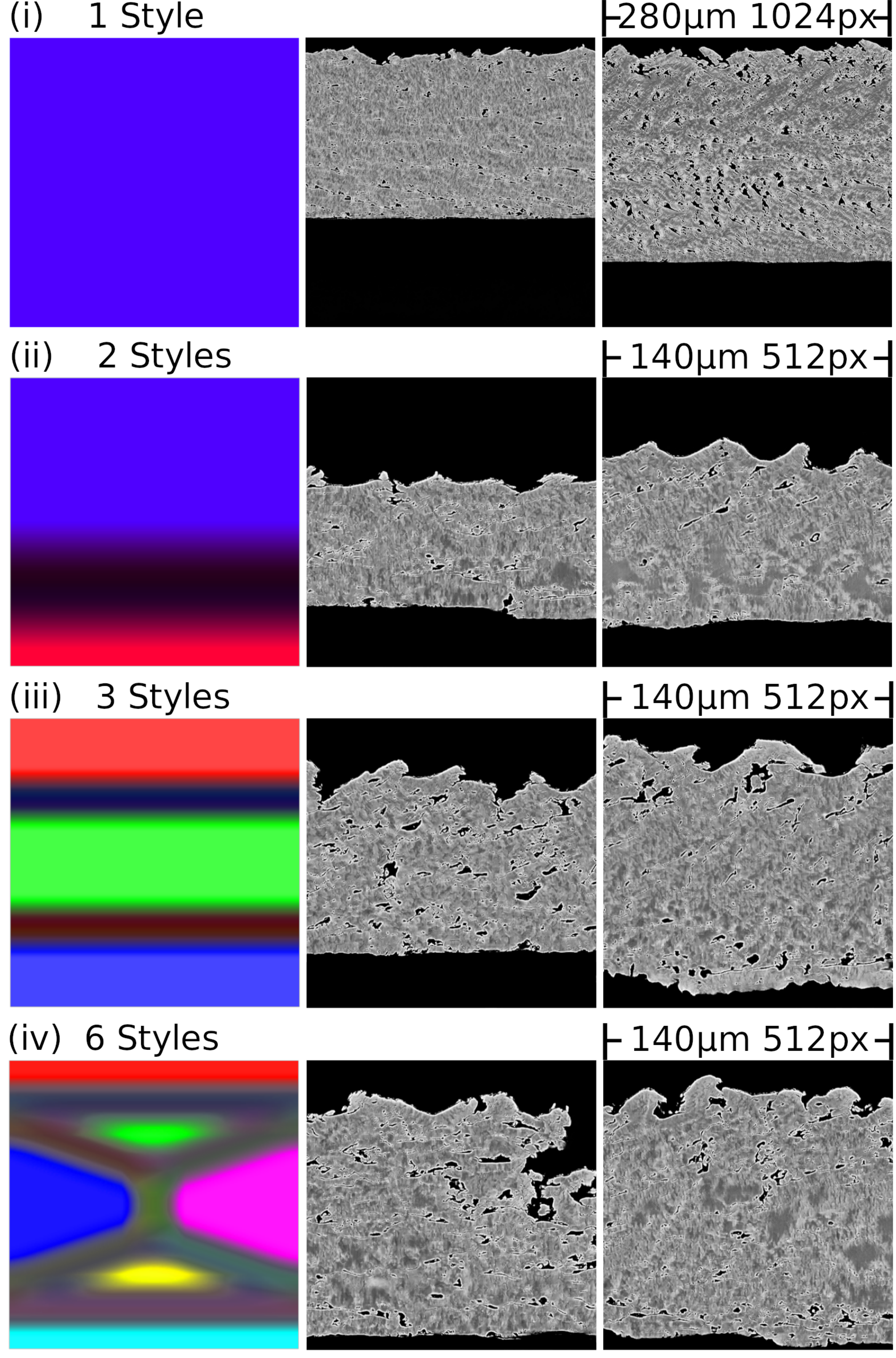}
\caption{{\bf Representative input maps and output images of the FluxGAN model.} The first column shows input maps and the second and the third columns show two representative images generated by the model for each input map, respectively. Increasing the number of styles and varying their spatial arrangement affects both the overall appearance and the local diversity of the generated images.}
\label{fig:generate_structures}
\end{center}
\end{figure}
Similar to InfinityGAN, the model uses 256 and 512 dimensions for local and global latent vectors, respectively, for unsupervised learning of structure and texture variations. We use coordinate grid to specify the location of the local structural patches to be sampled. We use sine functions with period $T$ = 56 to sample both $x$ and $y$ directions of the local latents. 
Our training dataset includes $64 \text{px} \times 64 \text{px}$ RGB image tiles, as mentioned previously. However, the model does not know the positioning of these $64 \text{px} \times 64 \text{px}$ tiles in the global image. To address the issue, the model uses smaller $37 \text{px} \times 37 \text{px}$ patches and learns to coordinate the patches relative to each other, during training. We use the same architecture of the discriminator as implemented in StyleGAN2. The schematic of the discriminator workflow is shown in Fig.~\ref{fig:diversity}(ii). The discriminator compares generated 37px $\times$ 37px patches with those included in the training set images. We assess the quality of images created by the model during training and find that the model achieved the smallest value of the Fr\'echet inception distance (FID) metric~\cite{heusel2017gans} of 2.64 after 660,000 training steps.

Figure~\ref{fig:diversity}(iii) shows representative output images that are generated based on single-style input maps. The input maps determine the different latent variables and drive the generation of the environments. Since the inputs are single-style maps, the model uses a single latent vector and generates all the regions of the microstructure. We vary the latent vector and compare the generated images, to examine the information encoded in the latent space. We obtain three sets of global latent vectors by using different random sampling from a unit Gaussian distribution. 
We use the three different latent vectors and keep all other variables (e.g., coordinates and styles) fixed, to obtain the three images shown in the first column of Fig.~\ref{fig:diversity}(iii). The structural environments of the images vary significantly when we change the global latent variables. The images show different regions of the coatings, such as the microstructure or the surface and air regions. Together the top surface, material, and bottom surface images constitute the large coating images. The images are representative of the regions marked with green, red, and blue boxes in Fig.~\ref{fig:diversity}(i). It is interesting to note that the the heat flux patterns across the images appear similar in magnitude, especially in regions away from the pores. For each choice of the global latent, we vary the styles and obtain the images shown in the second column of Fig.~\ref{fig:diversity}(iii). The structural environments remain very similar between the set of images shown in each row, however, the magnitudes of the heat flux are varied. This result implies that the global latent variables encode the environment (e.g., structure) and the styles are highly related to the physical phenomenon of interest (e.g., heat flux maps). This insight can offer great practical advantage. For example, styles can inform changes of heating conditions or reveal strategies to design structural features such that heat flux can be guided away from sensitive regions of the microstructures. Thus, FluxGAN model trained on style-based descriptions offers great practical advantage for designing microstructures that satisfy target thermal properties.

\begin{figure}
\begin{center}
\includegraphics[width=0.9\linewidth]{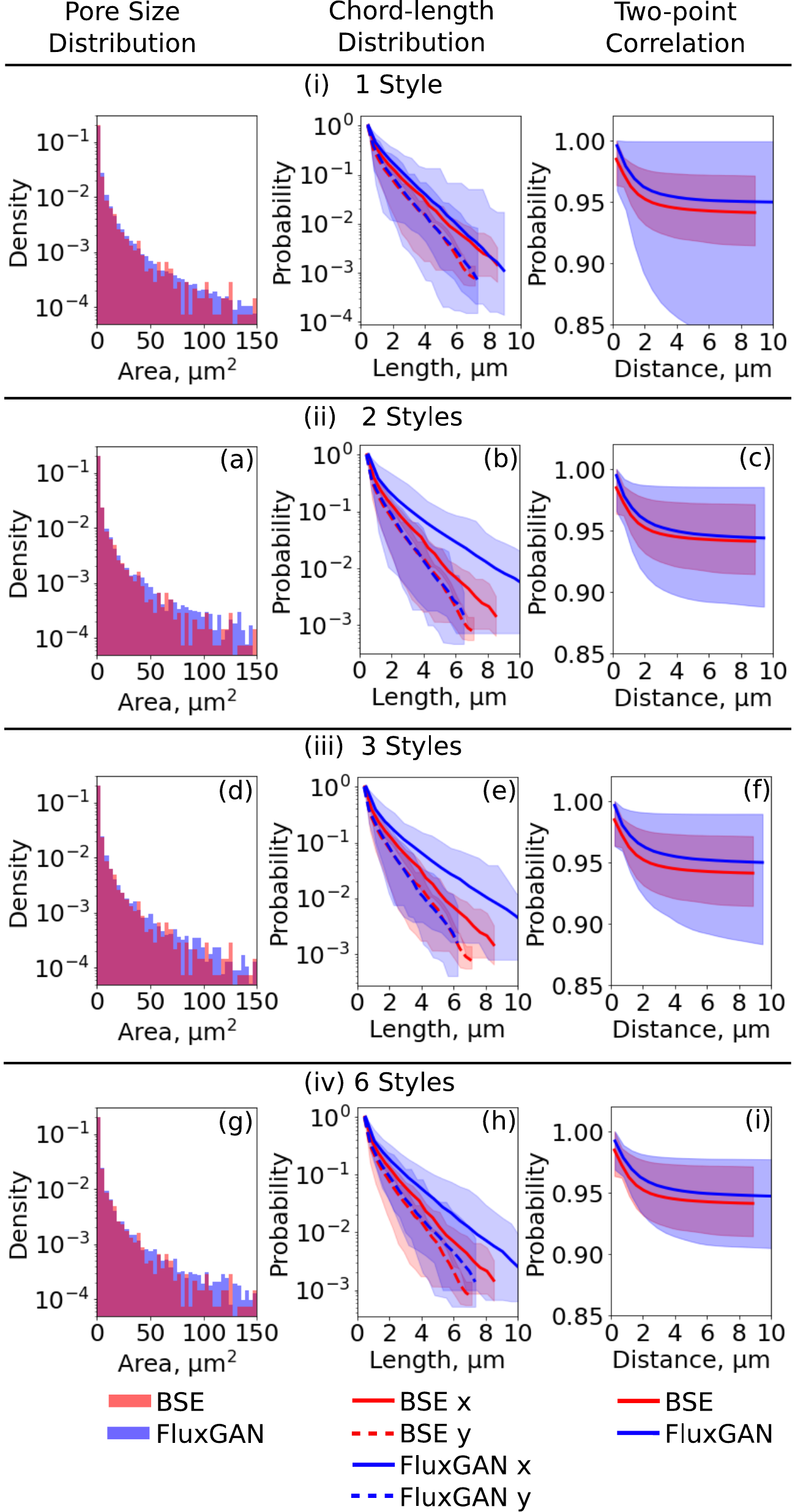}
\caption{{\bf Comparison of structural features of generated (blue) and experimental images (red).} We show the mean dependency (solid lines) and the range (shaded regions) for the chord-length distribution and the two-point correlation functions, in the second and the third column, respectively.} 
\label{fig:validate_structures}
\end{center}
\end{figure}

\section*{Generation Phase}
\subsection*{Unguided Generation}

Figure~\ref{fig:generate_structures} shows representative inputs (first column) and the corresponding images generated as output of the model (second and third column). The model input specifies the overall height and width of the image, the types and the number of styles, the spatial positioning of styles, and merging functions to obtain smooth interfaces between the regions. We show visual representations of the four input maps with different spatial arrangements of styles, in the first column of Fig.~\ref{fig:generate_structures}(i-iv). The input maps include one, two, three, and six randomly chosen styles, respectively. Note that the number of styles determines the global latent variables, and consequently, the overall appearance of the generated coating images, which is consistent with InfinityGAN spatial style fusion implementation. The images shown in the second and the third column of Fig.~\ref{fig:generate_structures}(i-iv) highlight the structural diversity in generated images for same input maps. We select 540 $1024 \text{px} \times 1024 \text{px}$ images generated with single-style and 50 $512 \text{px} \times 512 \text{px}$ images generated with each multi-style input map, and validate against experimental images using the procedure discussed in the following section. 
Although we show structural images in Fig.~\ref{fig:generate_structures}, the model outputs are RGB images that contain both structural and heat flux information. We only show the structural channel for the ease of discussion. 


\subsection*{Validation of Structural Information}

We compare the structural features of the generated images with those of the experimental BSE images. We analyze three features: pore size distribution, chord-length distribution, and  two-point correlation functions. Figure~\ref{fig:validate_structures}(i-iv) shows the comparison between the features of the experimental images and the images generated from one, two, three and six styles input maps, respectively. The first column shows the density of pores of given area, defined by the ratio between the number of pores with the specific area and the total number of pores. We calculate the pore sizes by counting the number of black pixels inside each pore. The presence of the black pores enclosed within the gray material region can be particularly noted from the experimental image shown in Fig.~\ref{fig:expimages}(ii). The smallest pore is of size 1px $\times$ 1px $\sim$ 0.07$\upmu$m$^2$. This size pores are the highest in number in the coatings as can be noted from Fig.~\ref{fig:validate_structures}(a,d,g). This could be a systematic artifact of the images. 
Overall, the density of different size pores in all four classes of images match with those in the experimental images. This implies that the model is able to produce the anisotropic pore distributions characteristic of the original thermal spray coatings shown in the BSE images. 
\begin{figure}
\begin{center}
\includegraphics[width=0.9\linewidth]{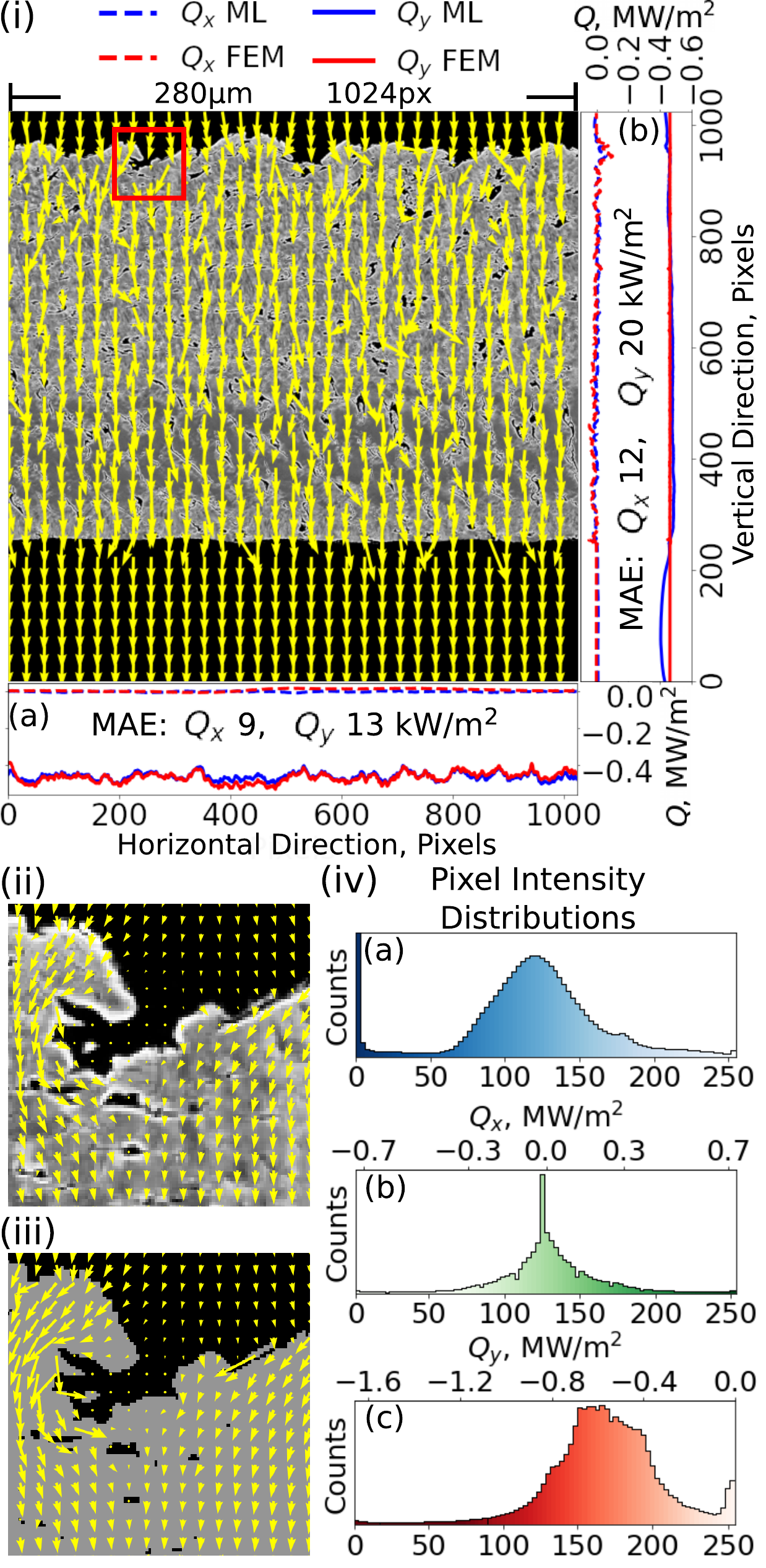}
\caption{{\bf Comparison of generated heat flux (blue) and FEM results (red).} (i) Full size generated image with single-style input map. Average generated heat flux (blue) along (a) horizontal and (b) and vertical directions, and compared to FEM results (red). Selected red boxed region is enlarged for comparison between (ii) generated and (iii) FEM results. (iv) (a) Structural and (b-c) heat flux pixel intensity distributions of 540 similarly generated images.}
\label{fig:validation}
\end{center}
\end{figure}
The porous coatings include numerous interfaces between the material phases and the pore phases. If an imaginary line is drawn in the porous medium, the line will intersect multiple material-pore interfaces. The lengths between these intersections are referred to as chord lengths in the respective phases. The chord length distribution function is then defined by the probability of finding a chord of given length in one of the phases~\cite{torquato1993chord, torquato2013random}. These functions provide a quantitative measure of the structure of the porous media. The second column of Fig.~\ref{fig:validate_structures} shows the comparison of the chord-length distributions in material phases of experimental (red lines) and generated images (blue lines). The solid and the dashed lines represent the chord-length distributions in the $x$ and $y$ direction, respectively.  
The results show that there is higher probability of finding longer chords in the $x$ direction compared to the $y$ direction in the images. This corresponds to the presence of layer-like features that can be noted from the experimental images, as shown in Fig.~\ref{fig:expimages}. The layer-like structures form along the vertical ($y$) direction when the droplets of melted material are sprayed in a regular interval during the thermal spray process. We show the comparison between the spatial two-point correlation functions in the third column of Fig.~\ref{fig:validate_structures}. The spatial correlation function between two points $\bf{r}_1$ and $\bf{r}_2$ is defined as $\langle f({\bf x}+{\bf r_1}) f({\bf x}+{\bf r_2}) \rangle$~\cite{berryman1986use, torquato2013random}, where and $\langle .. \rangle$ indicates a volume average over the spatial coordinate ${\bf x}=(x,y)$. The characteristic function $f(\bf{x})$ is either zero or one if the point lies in the material or the pore phase. We note a close match between the two-point correlation functions for all four cases. The result for the image generated with the six-style-input map is the closest to the one observed in BSE images. This can be attributed to the higher diversity of microstructures present in the images generated with six-style-input maps. 

\begin{figure}
\begin{center}
\includegraphics[width=0.9\linewidth]{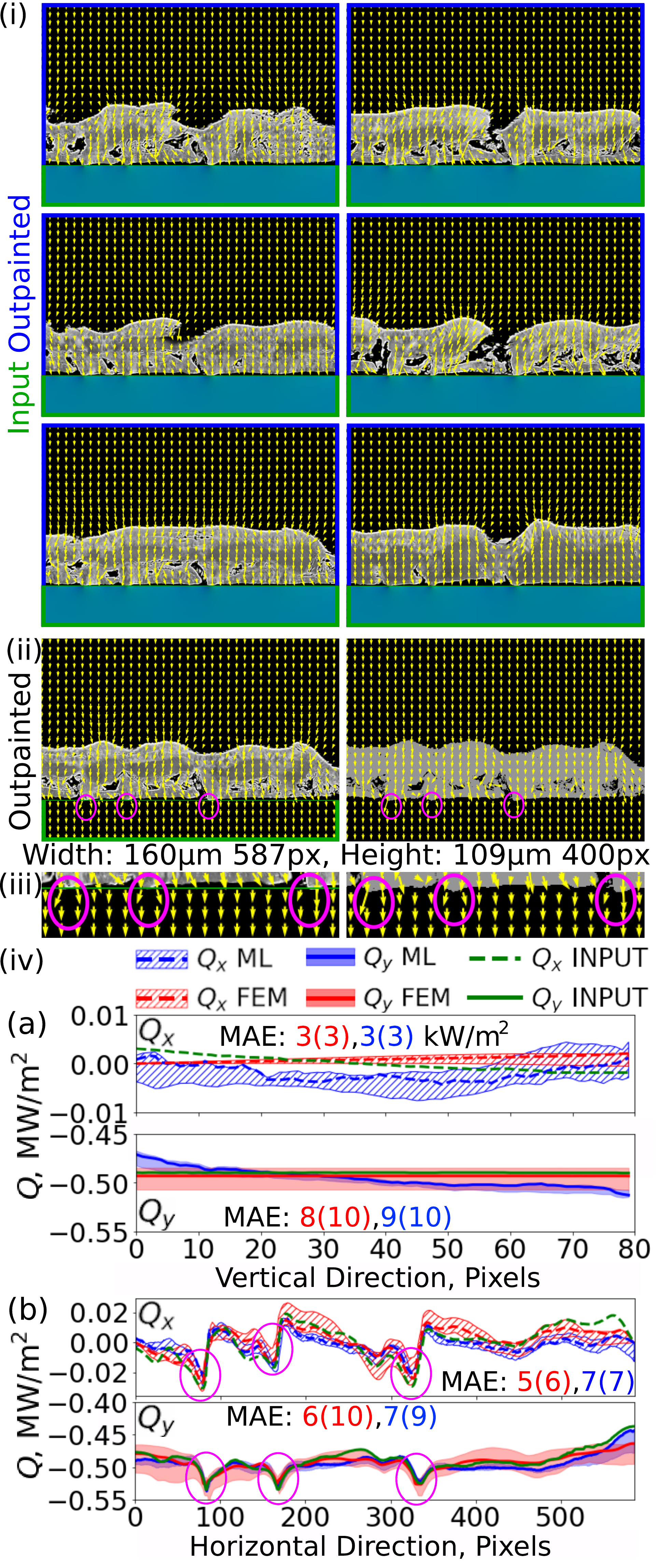}
\caption{{\bf Image outpainting with holistic heat flux distribution as input.} (i) Bottom air region (green box) is supplied as input. (ii) Outpainted image (left) and corresponding FEM results (right). (iii) Interface region of (ii). (iv) (a) Horizontal and (b) vertical averages of generated (blue) heat flux in input region, compared to input (green) and FEM results (red). MAEs for ML-INPUT (blue) and ML-FEM (red) comparisons reported for microstructure shown in (ii). MAEs averaged over seven systems are shown in parentheses}
\label{fig:outpaint_all}
\end{center}
\end{figure}

\subsection*{Heat Flux Data Validation}

We validate the heat flux of the generated coatings against values computed with FEM for the same microstructures. It is essential that the structural environments and the applied heating conditions of the FEM models are similar to those of the generated coatings. As we illustrate in the above sections, we generate coatings with input maps that include spatial arrangement of random choice of styles. The styles are closely associated with the heat flux patterns in different regions of the coatings. The coatings generated with different input maps not only show highly variable heat flux in the microstructures but also different heating conditions in the surface regions. The heating conditions in coatings generated with arbitrarily chosen input maps may not be consistent with the fixed heating conditions imposed in the FEM models. As we discuss in the ``Computation of Thermal Properties" section, we simulate a temperature bias of 400K to compute heat flux in all the BSE masks using FEM. To maintain similar heating conditions in generated structures and the FEM models, we provide only single-style-maps as input and generate coating images. We find that the trained FluxGAN model is able to generate full coating images that include material and air regions from sufficiently large ($1024 \text{px} \times 1024 \text{px}$) single-style-maps. We show an example generated microstructure and the associated heat flux in Fig.~\ref{fig:validation}(i). The yellow vector arrows represent the resultant of the horizontal ($Q_x$) and the vertical ($Q_y$) heat flux components. We produce masked images from the generated structural channel and compute the heat flux through the masked images using FEM calculations. We compute the average heat flux along the horizontal and the vertical directions of the coatings from the $Q_x$ and $Q_y$ values of the 1024 pixels of the square image. We show the comparison between the average generated (blue) and FEM-obtained (red) heat flux along the horizontal and the vertical directions, in Fig.~\ref{fig:validation}(i)(a) and (b), respectively. The generated heat flux is in good agreement with FEM results, with mean absolute errors (MAE) of 0.009-0.012 MW/m$^2$ for $Q_x$ and 0.013-0.020 MW/m$^2$ for $Q_y$ components, respectively. Note that the average $Q_x$ is approximately zero and the average $Q_y$ is always negative. The $Q_y$ components have a dominant contribution to the overall heat flux, since we assume that the temperature gradient is applied along top-to-bottom (negative $y$) direction. We show an enlarged image of the $100 \text{px} \times 100 \text{px}$ red boxed section within the large $1024 \text{px} \times 1024 \text{px}$ in Fig.~\ref{fig:validation} (ii). The corresponding masked image and the FEM result in shown in Fig.~\ref{fig:validation} (iii). The close match between the heat flux patterns can be particularly noted from the two images. We analyze 540 $1024 \text{px} \times 1024 \text{px}$ images generated with single-style-input maps. We show the pixel intensity distributions for the three channels in these 540 generated images in Fig.~\ref{fig:validation}(iv). The distributions show high diversity and are consistent with those in the training set, see Fig.~\ref{fig:distributions}(b-d).


\subsection*{Guided Generation}

Here, we illustrate how our FluxGAN model provides an expedited approach to design new coatings with target thermal properties. In the conventional approach for designing new coatings, one typically fabricates the coatings and then estimates the physical properties using experiments or computational modelling. The process is repeated till the fabricated microstructures satisfy target physical properties, as we outline in Fig.~\ref{fig:workflow}(ii). In contrast, the FluxGAN model directly guides the inverse design of coatings that satisfy target thermal properties. Given a target thermal property, the model generates rational designs of new microstructures that satisfy the properties. We demonstrate the targeted generation approach in Fig.~\ref{fig:outpaint_all}. We provide the model with the description of physical phenomena in a particular region, e.g., the heat flux near the heat sink, indicated by the green boxes in the images of Fig.~\ref{fig:outpaint_all}(i). The model input is a single 80px $\times$ 587px RGB image that we extract from one of the 410px $\times$ 587px images. The input image only contain information about the heat flux in the boxed air region where no material is present, referred to as $Q$ INPUT. The model uses outpaintaing technique to generate the microstructure, the top air region, and the associated heat flux ($Q$ ML) throughout the large 400px $\times$ 587px size images. We show the resulting images in Fig.~\ref{fig:outpaint_all}(i-ii). The model uses GAN inversion~\cite{creswell2018inverting,ma2018invertibility,abdal2019image2stylegan,bau2020semantic,zhu2020domain,cheng2022inout} to find the latent variables that match the input image and conditional coordination to independently generate spatially consistent micro-patches~\cite{lin2019coco}. Note that the model is capable of generating different coating images shown in Fig.~\ref{fig:outpaint_all}(i-ii), when supplied with the same input image. This is possible by selecting different latent variables during outpainting procedure.  

\begin{figure}
\begin{center}
\includegraphics[width=0.9\linewidth]{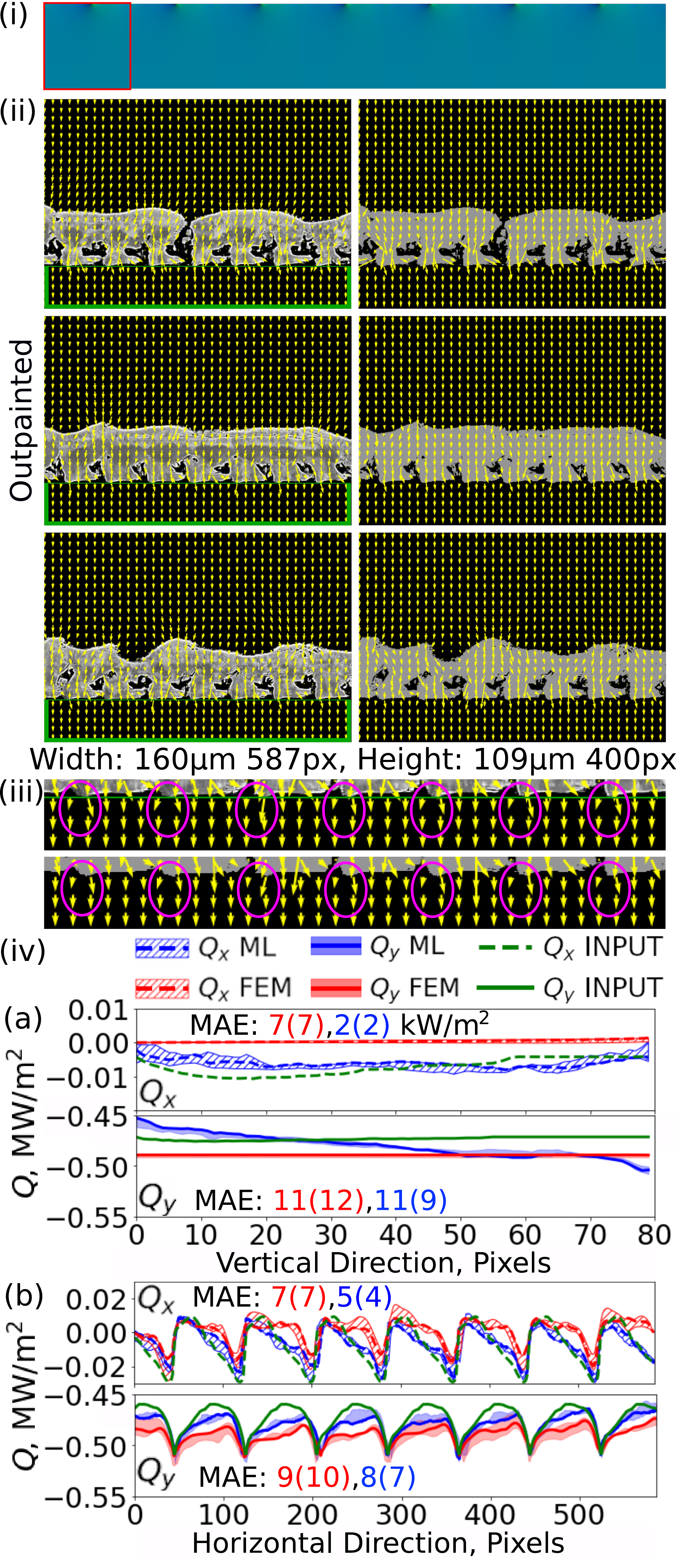}
\caption{{\bf Image outpainting with artificially constructed heat flux distribution as input.} (i) Input image. (ii) Example outpainted images (left) and corresponding FEM results (right). (iii) Interface region of representative image shown in (ii). (iv) (a) Horizontal and (b) vertical averages of generated (blue) heat flux in input region, compared to input (green) and FEM results (red). MAEs for ML-INPUT (blue) and ML-FEM (red) comparisons reported for bottom microstructure of (ii). Average MAEs of the three systems are shown in parentheses.}
\label{fig:outpaint_periodic}
\end{center}
\end{figure}

Finally, we calculate the heat flux associated with the generated microstructures using FEM, $Q$ FEM, for the same temperature bias of 400 K. We show an outpainted image and the corresponding FEM results for the masked image in the left and the right columns of Fig.~\ref{fig:outpaint_all}(ii), respectively. We show the comparison between $Q$ INPUT (green), $Q$ ML (blue) and $Q$ FEM (red) in Fig.~\ref{fig:outpaint_all}(iv). Figure~\ref{fig:outpaint_all} (iv)(a) and (b) show the average horizontal and vertical heat flux, computed from the $Q_x$ and $Q_y$ values of the 80 and 587 pixels of the input region, respectively. The shaded regions represent the range of heat flux values observed in all seven generated images. The MAEs represent the average absolute distance between the two plotted dependencies. The FluxGAN generated heat flux match with $Q$ INPUT and FEM results, with great accuracy. The errors range from 0.003 to 0.01 MW/m$^2$ for all the results, or around 2\% of the average heat flux value in the input region. The $Q_y$ components have a dominant contribution to the overall heat flux, compared to $Q_x$. This is expected since we assume that the temperature gradient is applied along top-to-bottom ($y$) direction. The $Q_y$ ML values show larger deviation at the boundaries since the FluxGAN model is unaware of insulating boundary conditions while the FEM results are obtained using this restriction. We marked three points of interest in the interface region with magenta circles, as shown in Fig.~\ref{fig:outpaint_all}(iii). These points represent the location of the cusps at the bottom material-air interface and are next to dips at the surfaces. They are closest to the heat sink, as can be noted by following the thin green line that marks the top boundary of the input region. The heat flux arrows at this locations (Fig.~\ref{fig:outpaint_all}(iii)) indicate that the heat is being guided sideways from the cusps, with the magnitudes being the highest among all values in the bottom air region (Fig.~\ref{fig:outpaint_all} (iv)(b)). Such characteristic heat flux patterns are associated with the different color environments around these points in all the input green boxed regions of Fig.~\ref{fig:outpaint_all}(i). Consequently, similar cusp points can be noted from all generated microstructures of Fig.~\ref{fig:outpaint_all}(i) and (ii). The common presence of these points in all generated microstructures refer to the fact that they all are rational designs that satisfy the target input heat flux. 

To further challenge the FluxGAN model, we test its ability to generate images for an artificially constructed input. We construct the input image by periodically extending the red-boxed 80px $\times$ 80px region marked in the top panel of Fig.~\ref{fig:outpaint_periodic}(i). The full panel shows the resulting input image of size 80px $\times$ 587px. We show three outpainted images and the corresponding FEM results in Fig.~\ref{fig:outpaint_periodic}(ii). We show the comparison of the average horizontal and vertical heat flux, computed from the $Q_x$ and $Q_y$ values of the 80 and 587 pixels of the input region, in Fig.~\ref{fig:outpaint_periodic}(iv)(a) and (b), respectively. The shaded regions represent the range of heat flux values observed in three different images. The periodic nature of heat flux can be particularly visible from the average vertical heat flux, shown in Fig.~\ref{fig:outpaint_periodic}(iv)(b). The FluxGAN generated heat flux match with $Q$ INPUT and FEM results, with great accuracy. The MAEs of the FluxGAN results range from 0.002 to 0.011 MW/m$^2$ when compared with the INPUT and from 0.007 to 0.012 MW/m$^2$ when compared with the FEM results, respectively. Average vertical heat flux agrees with the FEM results in the input region within 3\% error margin. These results demonstrate that the model is able to generate rational microstructures that best fit the criterion imposed by the input image. Note that the generated microstructures are highly diverse away from the periodic input region and do not show any visible repetitions. The diversity of generated images indicates that the model provides multiple designs of microstructures to satisfy target conditions. 

\subsection*{Extension to Three-Dimensional Domain}

\begin{figure*}
\begin{center}
\includegraphics[width=0.9\linewidth]{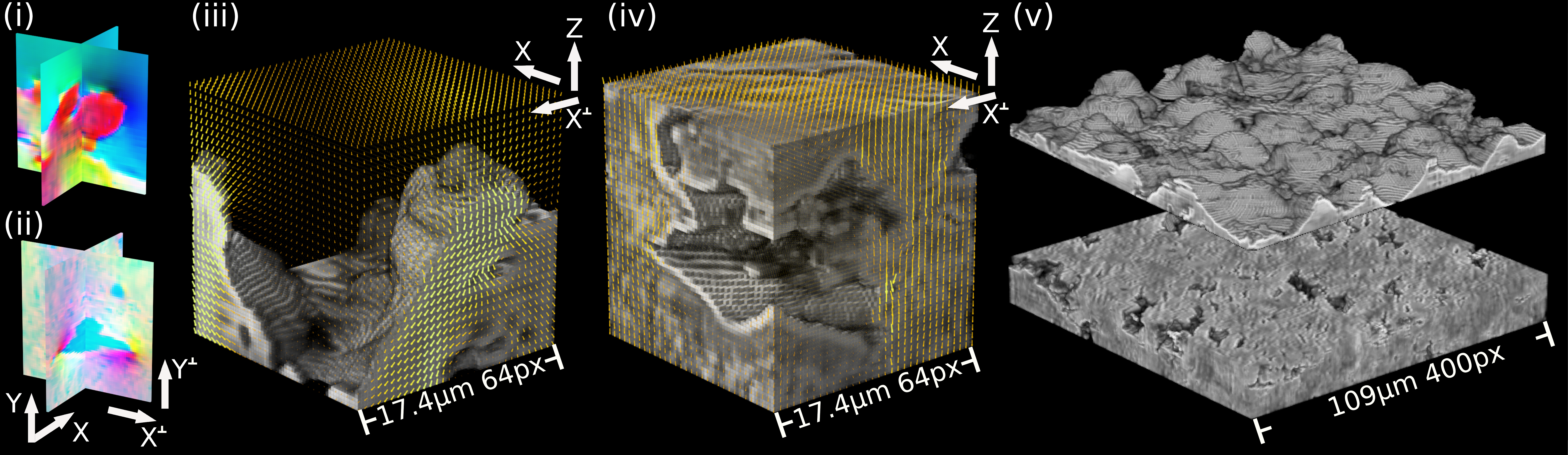}
\caption{{\bf Microstructure representation in 3D domain.} Two discriminators are implemented to test perpendicular RGB image slices for (i) surface and (ii) material regions. Representative volumes generated for (iii) surface and (iv) material regions with superimposed heat flux. (v) Two representative 3D microstructures generated by combined FluxGAN-SliceGAN approach.}
\label{fig:3D}
\end{center}
\end{figure*}


Although our study primarily focuses on 2D microstructures, here we discuss how we can leverage the FluxGAN model to generate 3D microstructures with high structural diversity. GANs can be trained to generate 3D volumes, which are useful for design optimization and detailed analysis of structural features and physical processes~\cite{kench2021generating, hsu2021microstructure, henkes2022three}. These models can be used for rapid generation of complex 3D structures over conventional approaches. A recent article used a bounded random walk algorithm to generate an interface, followed by seeding of random size particles to produce a two-phase material, resulting in large 128px $\times$ 10000px images. The images were used to train SliceGAN~\cite{kench2021generating} algorithm and output 3D configurations. However, the lack of diversity in such approach allows the generator to trick the discriminators, resulting in unrealistic 3D representations. We demonstrate that this issue can be resolved by training the GAN models with images that include both structural and physical phenomena information. We propose a combined approach that implements both the FluxGAN and the SliceGAN algorithms to generate complex 3D microstructures and associated heat flux environments with high diversity. The SliceGAN approach generates a volume by assimilating 2D slices, each of which is refined to match the training set images. However, a large number of diverse 2D images is required for successful training, especially to generate complex, anisotropic 3D materials with high density of interfaces. We leverage the FluxGAN model to generate two large 64px $\times$ 320000px images (17.4$\upmu$m $\times$ 8.7cm coatings) for the surface and the material regions. We make random selections of 64px $\times$ 64px patches from the generated images and pass them to two perpendicular discriminators of SliceGAN. Figure~\ref{fig:3D}(i) and (ii) show two perpendicular RGB slices for the surface and the material regions, respectively. Note that the heat flux components (red and green channels) are continuous across the two perpendicular slices and the blue channel produces matching structure in both the perpendicular planes. 

The objective of the generator of the SliceGAN model is to generate realistic volumes in which the 2D slices along $X$ and $Y$ directions pass the discriminator test. Note that the outputs from the generator include information in five channels: [BSE, $Q_X$, $Q_Y$, $Q_X^{\perp}$, $Q_Y^{\perp}$]. The first channel contains structural information as learned from the experimental BSE images. The other two sets of channels include heat flux data. ($Q_X$, $Q_Y$) and ($Q_X^{\perp}$, $Q_Y^{\perp}$) refer to the heat flux components of the two perpendicular slices, represented in graphical domain. The generated information is passed to the two discriminators. Each of the two perpendicular discriminators receive three-channel RGB images and share a common structural channel (BSE). However, each discriminator receives a separate set of heat flux data, ($Q_X$, $Q_Y$) or ($Q_X^{\perp}$, $Q_Y^{\perp}$). The model assimilates 64 slices in each of two perpendicular directions to produce 64 $\times$ 64 $\times$ 64 voxel volumes, as shown in Fig.~\ref{fig:3D}(iii) and (iv). For the generated 3D microstructures, we obtain the heat flux along the $Z$-direction, $Q_Z$, by averaging the vertical heat flux components $Q_Y$ and $Q_Y^{\perp}$ obtained from both discriminators. The resulting volumes have a four-channel representation: [BSE, $Q_X$, $Q_X^{\perp}$, $Q_Z$]. Finally, the SliceGAN model outputs two representative 3D configurations of size 64 $\times$ 400 $\times$ 400 voxels using the generated volumes, as shown in Fig.~\ref{fig:3D}(v). The resultant microstructures shows great structural diversity in both the surface landscape as well as the cross section of inner material region, which are not accessible in the experimental and training images. The combined FluxGAN-SliceGAN approach shows remarkable ability to infer 3D representation of microstructure and associated physical phenomenon from a single 2D image. We attribute the high quality of the representation to both large size of the training image and the heat flux data included in the image channels. 

\section*{Summary and Outlook}

We propose a generative ML model, FluxGAN, capable of generating arbitrary large images of microstructures and associated heat flux data. The model generates the information about heat flux that will be observed within the microstructures and the outside heat source or sink regions, if temperature gradients are applied externally. We train the model with RGB images in which the different color channels contain graphical domain representation of structural information from experimental images and heat flux data obtained from FEM calculations. In these images, the information about the structural features are superimposed with heat flux patterns that correspond to the structural features. During training, the model learns about the different combinations of structural features and heat flux patterns, referred to as styles, in an unsupervised manner. Once trained, the model generates large images for given inputs maps that specify spatial distribution of styles to be included in the output images. The generated microstructures show high diversity and include structural features that have characteristics similar to those present in the experimental images. We compare the generated heat flux with the results from separate FEM analyses on the corresponding generated microstructures. The FluxGAN generated heat flux show good agreement with FEM results. Furthermore, we demonstrate that the image generation can be guided by providing partial information to the model. We provide the model with the heat flux near the bottom heat sink region. The model shows remarkable ability to generate large microstructures, heat source regions and associated heat flux that are consistent with the input. The generated heat flux values lie within 3\% of those obtained with separate FEM simulations. Thus, our FluxGAN model offers an expedited approach for inverse design of microstructures with target thermal properties. The images show high structural diversity as well. We illustrate that the FluxGAN generated images can be leveraged to produce complex 3D microstructures, in combination with the SliceGAN architecture. We generate 0.1 meter long continuous and structurally diverse 2D structural images and their associated heat flux environment, to construct the 3D representations. 

In conclusion, FluxGAN offers fast, inexpensive, and reliable approach for design and analysis of 2D and 3D microstructures. The model is capable of generating arbitrary large environments with constant computation cost. This aspect is particularly advantageous over the FEM approach where memory usage scales linearly with the number of degrees of freedom. Our approach is general and can be extended to any continuous physical phenomena represented in graphical domain. The ability of the model to render arbitrary large environments while trained on small size patches offers great practical advantage. For example, the model will be highly effective in providing description of physical phenomena at the micro-meso or even macro scale, or guiding characterization and fabrication of complex devices, such as microelectronic chips with extreme design points.





\section*{Data availability}
The authors declare that the data supporting the findings of this study are available within the paper and its Supplementary Information files.

\section*{Acknowledgements}
We are indebted to Dr. M. R. Kracum for providing us with the two-dimensional backscatter electron images of the niobium thermal sprayed coatings, sharing valuable insights regarding the outstanding challenges of thermal spray coatings, and several helpful discussions. We gratefully acknowledge funding from the research partnership initiative between the College of Engineering \& Applied Science at the University of Colorado Boulder and the DOE-Sandia National Laboratories, Albuquerque, NM. This work utilized the Summit supercomputer, which is supported by the National Science Foundation (awards ACI-1532235 and ACI-1532236), the University of Colorado Boulder, and Colorado State University. The Summit supercomputer is a joint effort of the University of Colorado Boulder and Colorado State University.

\section*{Author contributions}

\noindent A.K.P and M.S. contributed to the acquisition and the analysis of data and the creation of new scripts used in the study. S.N. contributed to the conception and the design of the work, the interpretation of data, drafting and revision of the article.

\section*{Competing interests}

\noindent The authors declare no competing interests.

\bibliography{MLLiterature}

\section*{Supplementary Materials}
\begin{figure*}
\begin{center}
\includegraphics[width=0.9\linewidth]{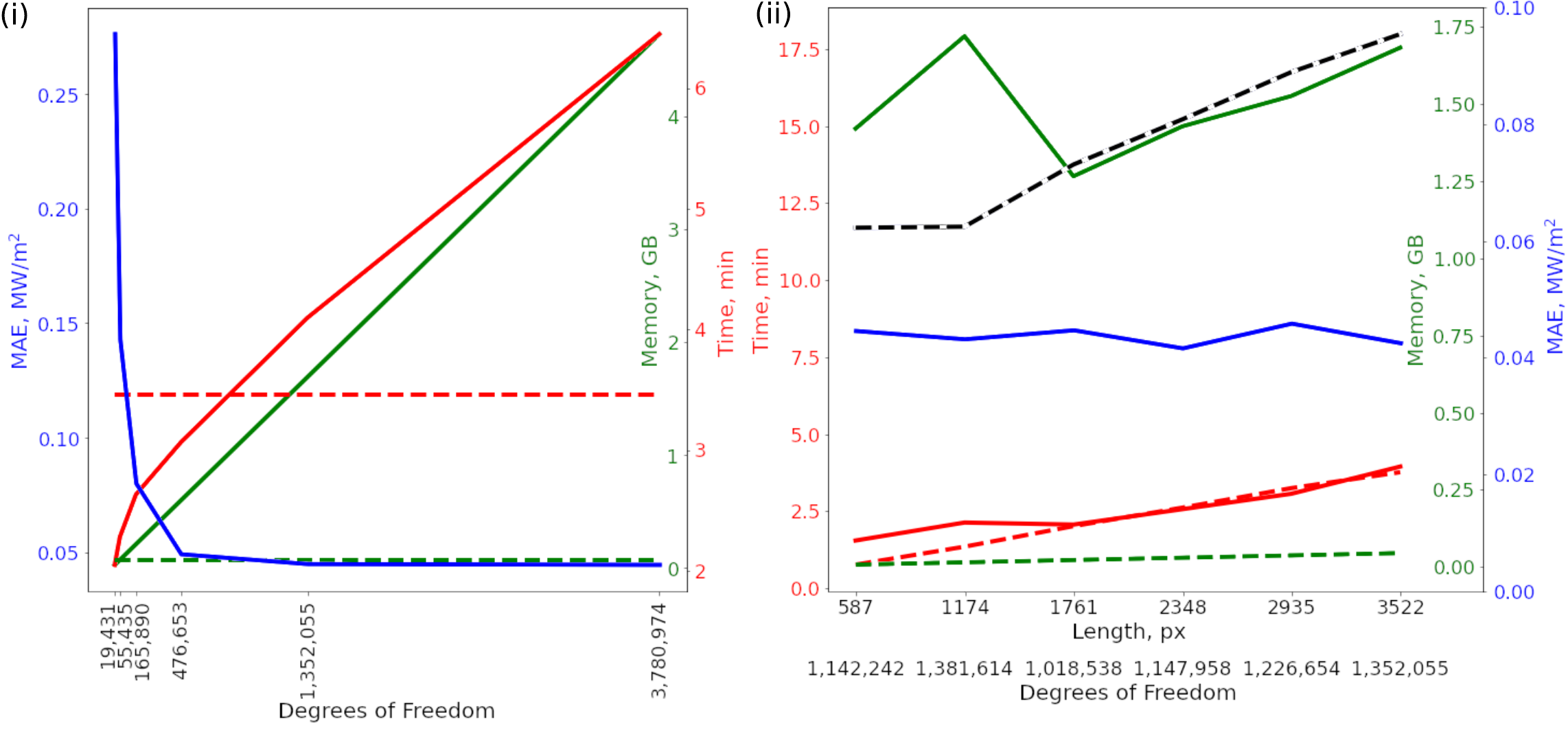}
\caption{{\bf Supplementary Materials} Computation and generation costs. (i) MAE and computational costs as a function of number of degrees of freedom for 400px $\times$ 3522px image. MAEs are calculated for the magnitude of heat flux. For the time (red) and memory (green) axes solid lines correspond to FEM, dashed lines correspond to ML. (ii) MAE and computational costs as a function of image length. Color lines show time (red) and memory (green) costs of FEM validation (solid) and ML generation (dashed). MAEs for heat flux magnitude are shown in blue. Latent inversion time cost is shown in black dashed line.}
\label{fig:timing}
\end{center}
\end{figure*}

Here we discuss and compare computational costs associated with ML model generation and FEM validation. We generate a set of images using periodically extended 80$\times$587 pixels input image discussed in section Guided Generation. We build a progression of input images extending the length by up to 6 times. The resulting outpainted images have the sizes 400px $\times$ (N$\times$587)px, with N=1,2,3,4,5,6. We demonstrate convergence of FEM validation mean absolute error (MAE) for the 400px $\times$ 3522px outpainted image in Fig.~\ref{fig:timing}(i). We show MAE for the total heat flux magnitude, namely, $Q=\sqrt{Q_x^2+Q_y^2}$, \\
MAE=$\sum_{i=1}^{H\times W}|Q_i^{ML}-Q_i^{FEM}|/(H\times W)$,\\
where $Q_i$ is the total heat flux magnitude at a given pixel, $H$ and $W$ are the height and width of an image. MAE converges quickly with the number of degrees of freedom (DoF) used for FEM calculation and saturates bellow 0.05MW/m$^2$ after 1 million DoF. For the further calculations we fix DoF around that value.
We also show calculation time (red solid line) and memory usage (green solid line) for FEM validation step. We observe their linear scaling with DoF. For comparison, the generation time (red dashed line) and memory usage (green dashed line) are also shown for ML model.  In Fig.~\ref{fig:timing}(ii) we compare ML generation and FEM calculation costs. Both FEM (red solid line) and ML (red dashed line) have similar linear time scaling with image size for the DoF fixed around 1 million. However, the memory usage is significantly higher for FEM (green solid line) compared to ML (green dashed line). We also show that MAE (blue solid line) stays consistently small (bellow 0.05MW/m$^2$) for all image sizes. The highest time cost is associated with latent inversion (black dashed line). The low memory costs associated with ML are due to patch-by-path generation approach, where each patch (37px $\times$ 37px) is generated independently to produce large scale images. The time costs for ML generation shown here are for generating one patch at a time. This can be improved significantly since the model is capable of generating patches in batches, allowing for the highly parallelized output. 

\end{document}